\documentclass[3p,times]{elsarticle}

\usepackage{ecrc}

\volume{00}

\firstpage{1}

\journalname{Nuclear Physics A}

\runauth{F. Garibaldi et al.}

\jid{NUPHA}

\jnltitlelogo{Nuclear Physics A}

\CopyrightLine{2012}{Published by Elsevier Ltd.}

\usepackage{amssymb}

\usepackage[figuresright]{rotating}

\begin{document}

\begin{frontmatter}

\dochead{}



\title{High Resolution  Hypernuclear Spectroscopy
at Jefferson Lab Hall A}
\address[Roma]{INFN Sezione di Roma, Rome, Italy}
\address[ISS]{INFN gr. Sanit\`a coll. Sezione di Roma and Istituto
Superiore di Sanit\`a, Rome, Italy}
\address[Praha]{Nuclear Physics Institute, \v{R}e\v{z} near Prague,
Czech Republic}
\address[TUM]{Technische Universit\"{a}t M\"{u}nchen, Excellence Cluster Universe, 
85748 Garching, Germany}
\address[Bari]{INFN Sezione di Bari and Dipartimento di Fisica, Bari, Italy}
\address[Tre]{INFN Sezione di Roma Tre, Rome, Italy}
\address[JLab]{Thomas Jefferson National Accelerator Facility, Newport
News, VA 23606, USA}
\address[FIU]{Florida International University, Miami, FL 32306, USA}
\address[BNL]{Brookhaven National Laboratory, Upton, NY 11973, USA}
\author[Roma,ISS]{F.~Garibaldi\corref{cor1}}
\ead{franco.garibaldi@iss.infn.it}
\author[Praha]{P.~Byd\v{z}ovsk\'y}
\author[ISS]{E.~Cisbani}
\author[TUM]{F.~Cusanno}
\author[Bari]{R.~De~Leo}
\author[ISS]{S.~Frullani}
\author[Tre]{M.~Iodice}
\author[JLab]{J.~J.~LeRose}
\author[FIU]{P.~Markowitz}
\author[BNL]{D.~J.~Millener}
\author[Roma]{G.~M.~Urciuoli}
\author{\\ for the Hall A Collaboration}
\cortext[cor1]{Corresponding author}
\begin{abstract}
The characteristics of the Jefferson Lab electron beam, together with those of the experimental equipment, offer a unique opportunity to study hypernuclear spectroscopy 
via electromagnetic induced reactions. Experiment 94-107 started a systematic study on 1p-shell targets, $^{12}C$, $^{9}Be$ and $^{16}O$. For $^{12}C$ for the first time 
measurable strength in the core-excited part of the spectrum between the ground state and the p state was shown in 
$^{12}_{\Lambda}B$ spectrum. A high-quality $^{16}_{\Lambda}N$ spectrum was produced for the first time with sub-MeV energy resolution. A very precise $\Lambda$ binding energy value for $^{16}_{\Lambda}N$, calibrated against the elementary $(e,e'K^+)$ reaction on hydrogen, has also been obtained. 
$^{9}_{\Lambda}Li$ spectrum shows some disagreement in strength for the second and third doublet with respect to the theory. 
\end{abstract}
\end{frontmatter}
\section{Introduction}
\label{Intro}
Hypernuclear-multibaryonic systems with non-zero strangeness are an important branch of contemporary nuclear physics. The strange baryon is an impurity in the system that allows one to measure the system response to the stress imposed by it. The study of its propagation can reveal configurations or states not seen in other ways and gives interesting, important insight into the structure of ordinary nuclear matter. In fact, the hyperon is not affected by the Pauli principle and can penetrate deep inside the nucleus. More generally, the nucleus provides a unique laboratory for studying the lambda interaction. Hypernuclear experimental studies up to now have been carried out by hadron-induced reactions, ($K^-$, $\pi^-$) or ($\pi^+$, $K^+$), with limited energy resolution (about 1.5 MeV at the best). The experimental panorama can be greatly improved using electro-production of strangeness, characterized by large momentum transfer $q \gtrsim$ 350 MeV/c and strong spin-flip terms even at zero kaon production angles.  Photons may excite both natural and unnatural parity, low and high-spin hypernuclear states including states with a deeply-bound lambda hyperon. Moreover, in the case of ($K^-$, $\pi^-$) or ($\pi^+$, $K^+$), the elementary production of the lambda hyperon occurs on the neutron, while in the electromagnetic production, the ($K^+$, $\Lambda$) pair production occurs on the proton making it possible to study hypernuclei not otherwise available (e.g. $^{12}_{\Lambda}B$) including the hypernuclei with a large excess of neutral baryons ($^{7}_{\Lambda}He$, $^{9}_{\Lambda}Li$). Comparison of the spectra of mirror hypernuclei (e.g…) can then shed some light on the charge asymmetry of hyperon - nucleon forces. The disadvantage of smaller electromagnetic cross sections is partially compensated by the high current, high duty cycle, and high energy resolution capabilities of the beam at Jefferson Lab. Missing mass resolution as good as $\approx$ 600 KeV (FWHM) can be attained, in principle, assuming beam energy stability of 2.5x$10^{-5}$ and High Resolution Spectrometer momentum resolution of $10^{-4} $(FWHM). A large effort has been made by the collaboration to improve the experimental apparatus for this specific experiment. The Experiment E94-107, in Hall A at Jefferson Lab started a systematic study on 1p-shell targets, $^{12}C$, $^{9}Be$ and $^{16}O$ \cite{Garibaldi}
\section{Experimental Equipment}

In order to effectively do high-resolution hypernuclear spectroscopy experiments with electron beams three ingredients are required: (1) A high quality, high current, high duty factor,  4 GeV electron beam. (2) Two high resolution spectrometer arms. (3) Excellent particle identification. Hall A at JLab is well suited to perform $(e,e'K^+)$ experiments. Scattered electrons can be detected in the High Resolution Spectrometer (HRS) electron arm while coincident kaons are detected in the HRS hadron arm \cite {NIMA}. 
The experiment required a continuous 100~$\mu$A electron beam with very small energy spread and vertical spot size (energy spread $\sim$ 2.5x$10^{-5}$ spot size $\sim$ 100 $\mu$m). The energy spread was monitored continuously using a Syncrotron Light Interferometer (SLI) to monitor the physical size of the beam at a point in the Hall A arc with large dispersion. A 100~mg/cm$^2$ target was used for the $^{12}C$ experiment with an electron beam current of 100~$\mu$A. The strong inverse dependence of the cross section on $Q^2$, squared virtual photon 4-momentum transfer, calls for measurements at low $Q^2$. To maximize the cross section, the electron  scattering angle must be minimized, in compromise with the increasing background from processes at very forward electron angles. To minimize the momentum transferred to the hyperon, and maximize the cross section, a detection angle $\theta_K$  must be chosen near the virtual photon direction. The high beam energy results in a relatively high momentum for the kaon, as required to keep a reasonable survival fraction in the spectrometer (25~m flight path). So, kinematics were set to particle detection at 6$^{\circ}$ for both electrons  and kaons, incident beam energy of 3.77 GeV, scattered electron momentum of 1.56 GeV/c, and kaon momentum of 1.96~GeV/c. In order to allow experiments at very forward angles, a superconducting septum magnet was added to each HRS. Particles at scattering angles of 6$^{\circ}$ are deflected by the septum magnets into the HRS. The energy resolution depends on the momentum resolution of the HRS spectrometers, on the straggling and energy loss in the target, and on the beam energy spread. The high background level demands a very efficient PID system with unambiguous kaon identification. The standard PID system in the hadron arm is composed of two aerogel threshold Cherenkov counters ($n_1$ = 1.015 $n_2$ = 1.055) \cite{NIMA}. However, due to inefficiencies and delta-ray production, the identification of kaons has contamination from pions and protons. This has driven the design, construction, and installation of a Ring Imaging CHerenkov (RICH) detector, conceptually identical to the ALICE HMPID design, in the hadron HRS detector package. It uses a proximity focusing geometry, a CsI photocathode, and a 15 mm thick liquid perfluorohexane radiator. A detailed description of the layout and the performance of the RICH detector is given in \cite{Iodice}. 
\section{Results for $^{12}C(e,e'K^+)^{12}_{\Lambda}B$}
\label{Carb}
\begin{figure}[htb!]
\centering
\includegraphics[width=9.5cm, angle=0]{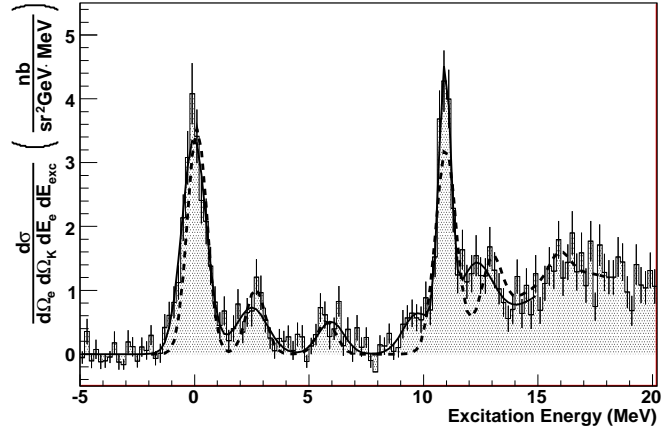}
\caption{Excitation energy spectrum of the
$^{12}C(e,e'K^+)^{12}_{\Lambda}B$ compared with a new theoretical
calculation (dashed line, see text).}
\label{fig:newcarb}
\end{figure}
The measured excitation energy spectrum of $^{12}C$ from a 100 mg/cm$^2$ target is shown in Fig.~\ref{fig:newcarb}. The results of the investigation of the $^{12}C(e,e'K^+)^{12}_{\Lambda}B$ have been already described elsewhere \cite{Mauro}. The theoretical cross sections, for all of the
investigated hypernuclei, are obtained in the framework of the
distorted wave impulse approximation (DWIA) \cite{DWIA} using the
Saclay-Lyon (SLA) model \cite{SLA} for the elementary
$p(e,e'K^+)\Lambda$ reaction. Shell-model wave functions are
determined using a parametrization of the $\Lambda N$ interaction that
fits the precise $\gamma$-ray hypernuclear spectra of
$^{7}_{\Lambda}Li$ \cite{Ukai}.
The agreement with the data for the ground state and the core-excited states is very good. For the first time a measurable strength with good energy resolution has been observed in the core-excited part of the spectrum. The $s_{\Lambda}$ part of the spectrum is well reproduced by the theory. The distribution of strength within several MeV on either side of the strong $p_{\Lambda}$ peak at 10.95 MeV should stimulate theoretical work to better understand the pΛ region. 
\section{Results for $^{16}O(e,e'K^+)^{16}_{\Lambda}N$}
\label{Oxy}
\begin{figure}[htb!]
\centering
\includegraphics[width=9.5cm, angle=0]{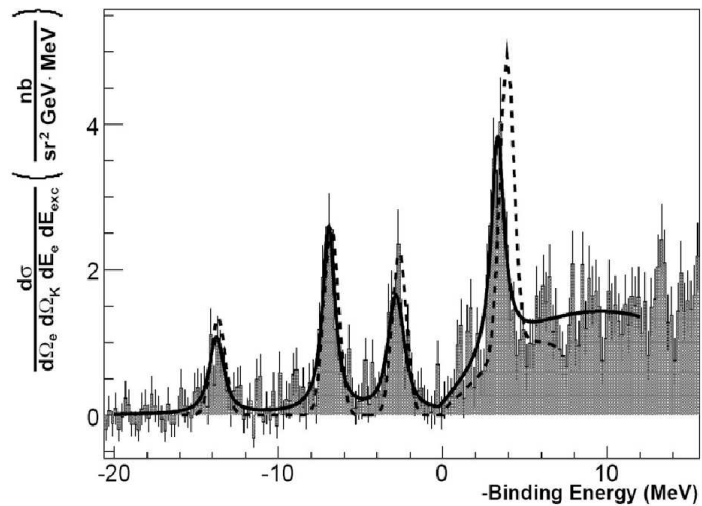}
\caption{Excitation energy spectrum of the
$^{16}O(e,e'K^+)^{16}_{\Lambda}N$ compared with the theoretical
calculation.}
\label{fig:oxy}
\end{figure}
The results of the investigation of the $^{16}O(e,e'K^+)^{16}_{\Lambda}N$ have been recently published elsewhere \cite{OxyPaper}. A waterfall target has been used. Kinematics were set to electron detection at  6$^{\circ}$ for scattered electrons with momentum of 1.44 GeV/c, incident beam energy of 3.66 GeV, virtual photon energy of 2.2 GeV with $Q^2$ = 0.06 GeV$^2$. Produced kaons were detected with momenta of 1.96~GeV/c at 6$^{\circ}$. Fig.~\ref{fig:oxy} shows the  cross-section for the $^{16}_{\Lambda}N$ hypernuclear spectrum produced on Oxygen nuclei at $\theta^{lab}_K$= 6$^{\circ}$. The dotted line is a result of the theoretical calculation obtained with the SLA model for the elementary cross section and by using J. Millener calculations for the hypernuclear structure. The continuous black line is the fit of the curve with same procedure used for the Carbon spectrum. The overall picture shows a very good agreement between the data and the calculations  in terms of positions and relative strength of the levels. The waterfall thickness used during the experiment is 75 mg/cm$^2$, which was measured from the cross-section of the elastic reaction on hydrogen. The measured $\Lambda$ binding energy  = 13.57 $\pm$ 0.25 MeV obtained for the first peak is an important quantity because there are few emulsion events for the heavier p-shell hypernuclei and these events tend to have ambiguous interpretations and the reactions involving the production of a $\Lambda$ from a neutron are more difficult to normalize. 
\section{Results for $^{9}Be(e,e'K^+)^{9}_{\Lambda}Li$}

\begin{figure}[htb!]
\centering
\includegraphics[width=12cm, angle=0]{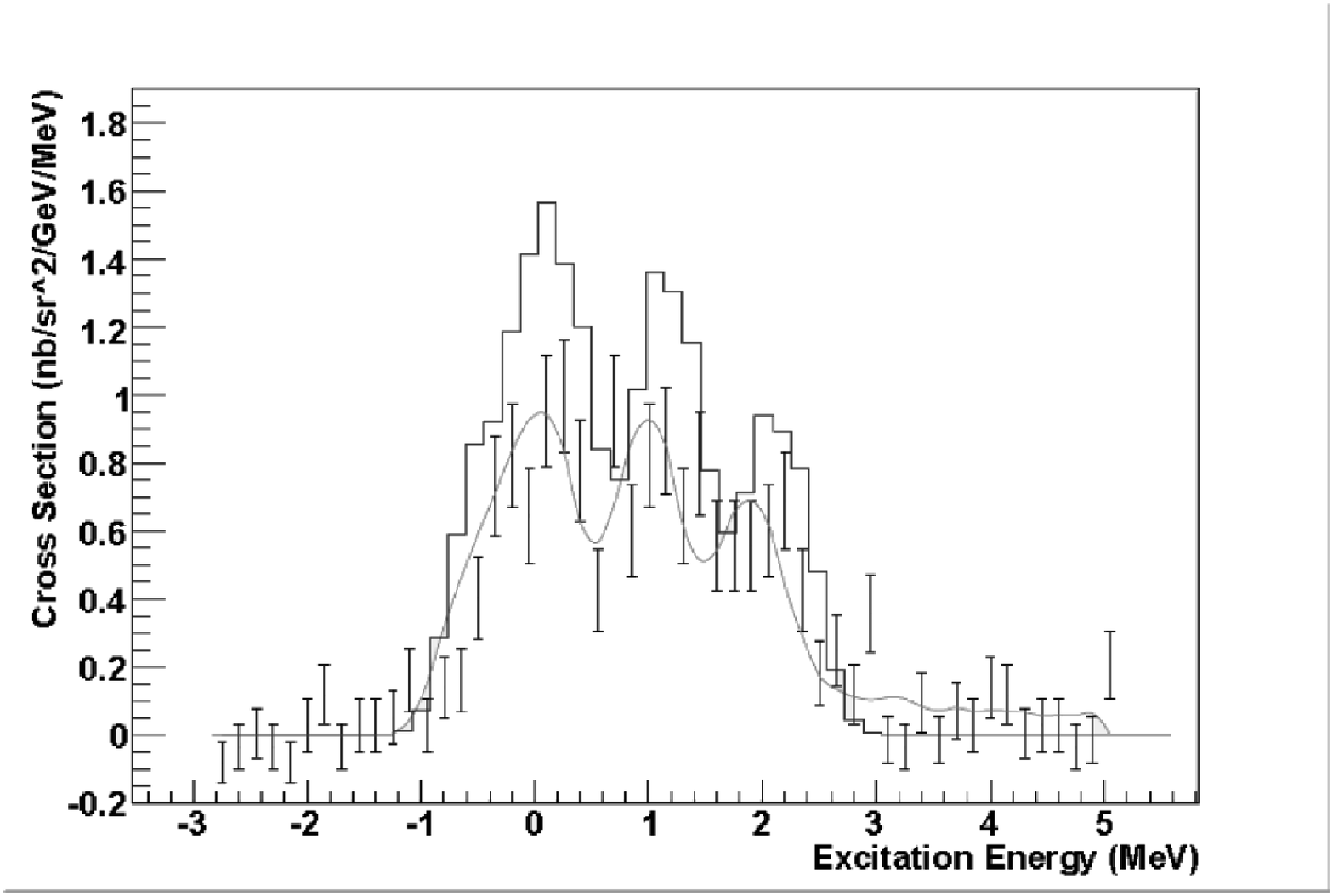}
\caption{\small \label{RadiativeCorr}
Correction of radiative effects on preliminary $^{9}_{\Lambda}Li$ 
excitation energy spectrum. The 
black points are the experimental data, the gray curve represents 
simulated data 
including radiative effects, the histogram represents simulated 
data with no radiative effects. See text for details. 
}
\end{figure}

\begin{figure}[htb!]
\centering
\includegraphics[width=8.5cm, angle=270, trim=0 0 0 0,clip]
{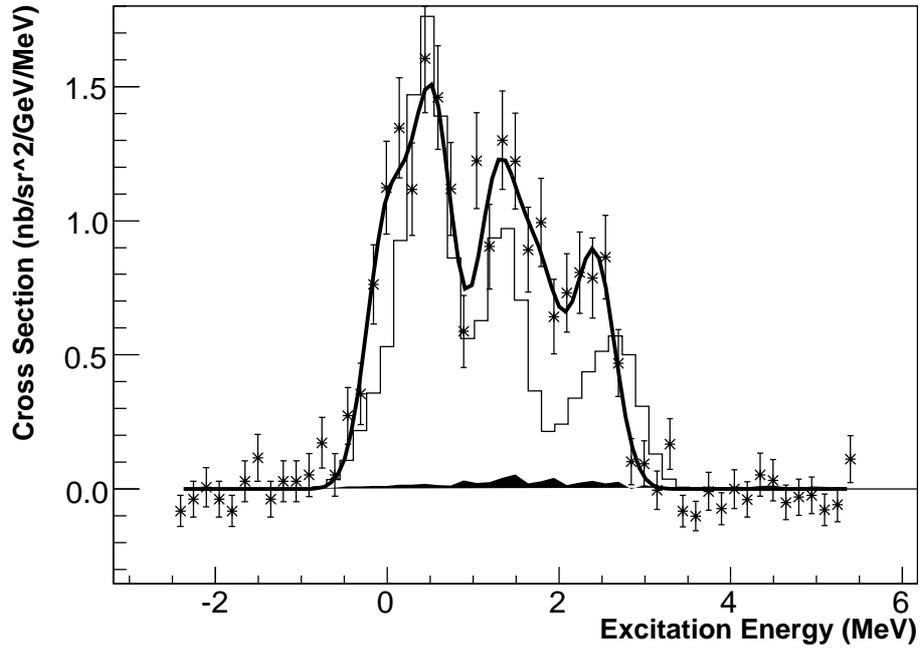}
\caption{\small \label{Fig2Label}
Five-peak gaussian fit on preliminary $^{9}_{\Lambda}Li$ excitation 
energy spectrum corrected for radiative effects. Data points 
are reported with their statistical errors, the black-filled 
histogram represents the systematical errors, the thin-black-line 
histogram represents the expectations from the theoretical model 
\cite{DWIA}, the thick black curve is a 
fit of the experimental data as explained in the text. 
}
\end{figure}

\begin{figure}[htb!]
\centering
\includegraphics[width=8.5cm, angle=270, 
trim = 0 0 0 0,clip]{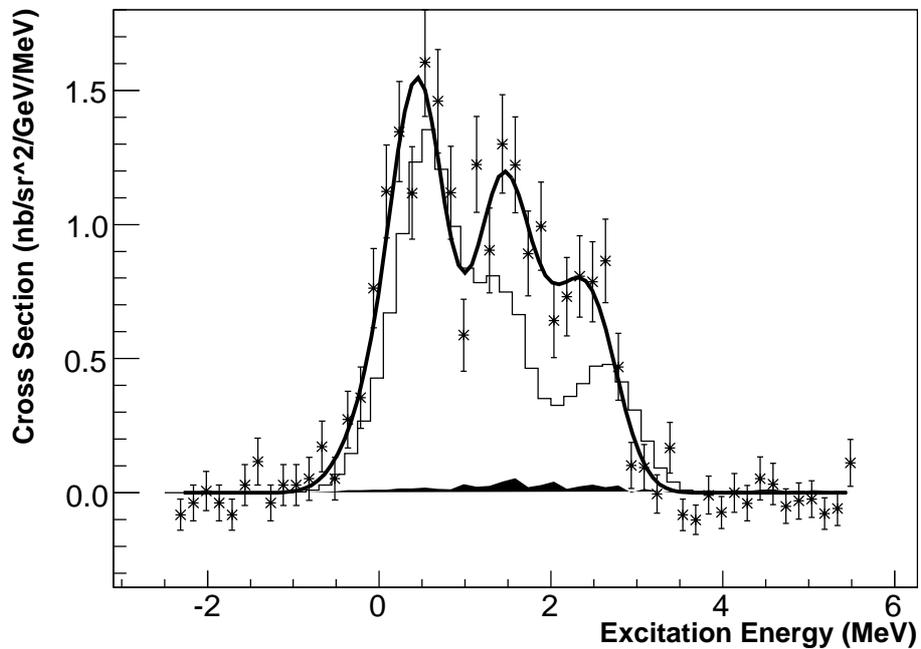}
\caption{\small \label{Fig3Label}
Same as Fig.~\ref{Fig2Label} in case of constraining the fit according 
to the theoretical model. See text for details. For the theoretical 
histogram the width of 760 keV (FWHM) was used contrary to 570 keV 
in Fig.~\ref{Fig2Label}.
}
\end{figure}

The analysis of $^{9}_{\Lambda}Li$ energy spectrum is 
not yet finalized: 
with respect to what was reported in \cite{HypX} a fine correction for 
radiative effects has been performed,
Fig.~\ref{RadiativeCorr} shows the effect of this correction.
The Monte Carlo code SIMC \cite{SIMC} has been used for this purpose: 
once the simulated 
data fit well the experimental data, then the radiative effects 
in SIMC are turned off and the ratios between simulated data with no 
radiative effects and simulated data including radiative effects are the 
bin-by-bin correction factors for the experimental points. 
\\
Fig.~\ref{Fig2Label} shows a preliminary 
$^{9}_{\Lambda}Li$ excitation energy spectrum corrected for the 
radiative effects. 
Since the DWIA calculations predict five states, a five-peak gaussian 
fit (thick black curve) is performed on the data 
points, with the only constraint of having the same width for the five peaks. 
The resulting width for the five peaks is 570~keV (FWHM), 
then the histogram of the predicted values is obtained assuming the 
same width for the theoretical expectations (thin line).  
\\
Fig.~\ref{Fig3Label} shows the same data as in Fig.~\ref{Fig2Label} 
but a different fit is calculated. In this case the five-peak 
gaussian fit is constrained by the model: the separation and the 
relative amplitude of the individual levels composing the first and 
the third peak (doublets) are fixed according to the theoretical 
expectations. In other words, this fit corresponds to a 
three-peak fit where the internal structure of the complex peaks is 
determined by the theory.
As in Fig.~\ref{Fig2Label}, the 
width of the five peaks is constrained to be a single value, resulting 
here in 760~keV (FWHM).\\ 
According to this preliminary analysis, the position of the peaks and 
the amplitude of the first doublet are in 
good agreement with the model, the amplitude of the excited 
states are instead underestimated by the theoretical predictions. 
The reason for the disagreemnt in strength for the second and third peak, is hard to ascertain and could be due to a number of deficiences in the structure or reaction calculations
\\  
\section{Conclusion}
\label{concl}
The systematic study of hypernuclear spectroscopy by electroproduction 
of strangeness performed at Jefferson Laboratory in Hall A has been very successful. It provided important  elements for a better understanding of the baryon-baryon interactions 
and production mechanism in strangeness physics. The experiment was successful but challenging because important modifications to the Hall A apparatus were needed. The new experimental equipment, aerogel threshold detectors, septum magnets and the RICH detector all gave excellent performance. Unprecedented energy resolution  and very clean, background free, spectra were  obtained.  The results of the hypernuclear spectroscopy performed on $^{12}C$, $^{16}O$ and $^{9}Be$ targets  provide important elements for a better understanding of strangeness physics.  Results from $^{12}C$ showed for the first time significant strength in the core excited part of the spectrum. The s-shell part of the spectrum was  well reproduced by the theory, while the p shell part does not. This will allow extraction of more information on the $\Lambda$ – N interaction. Moreover, for $^{16}O$, thanks to the calibration with the hydrogen present in the waterfall target, a very precise determination of the $\Lambda$ binding energy was obtained. In the case of $^{9}Be$ the measured cross sections are in good agreement for the first peak with the values predicted using the SLA model and simple shell model wave funnction. The reason for the disagreemnt in strength for the second and third peak, is hard to ascertain and could be due to a number of deficiences in the structure or reaction calculations.
\\
\\
\textbf{Acknowledgements}
\\
This work was supported by US Department of Energy contract
DE-AC05-84ER40150 Modification No. M175 under which formerly the
Southeastern Universities Research Association and presently the
Jefferson Science Associates LLC operates the Thomas Jefferson
National Accelerator Facility, by the Italian Istituto Nazionale di
Fisica Nucleare (INFN), by the US Department of Energy under contracts
W-31-109-ENG-38, DE-FG02-99ER41110, and DE-AC02-98-CH10886, by the
GACR Grant No. P203/12/2126, and by the French CEA
and CNRS/IN2P3. \\
\\
\textbf{References}

\begin{thebibliography}{99}
\bibitem{Garibaldi}  F. Garibaldi, S. Frulllani, J. LeRose and P. Markowitz spokesperson, Jlab Experiment 94-107, High resolution 1p shell Hypernuclear Spectroscopy. 
\bibitem{NIMA} J. Alcorn {\it et al.}, Nucl. Instrum. Methods Phys.
Res., Sect. A {\bf 522} (2004) 294.
\bibitem{Iodice} M. Iodice {\it et al.} Nucl. Instr. and Methods A 553, 231 (2005)
\bibitem{Mauro} M.~Iodice {\it et al.}, Phys. Rev. Lett. {\bf 99} (2007) 052501.
\bibitem{DWIA} M.~Sotona and S. Frullani, Prog. Theor. Phys. Suppl
{\bf 117} (1994) 151.
\bibitem{SLA} T. Mizutani, C. Fayard, G.-H. Lamot, and B. Saghai,
Phys. Rev. C {\bf 58} (1998) 75.
\bibitem{Ukai} M.~Ukai {\it et al.}, Phys. Rev. C {\bf 73} (2006) 012501(R).
\bibitem{OxyPaper} F.~Cusanno {\it et al.}, Phys. Rev. Lett. {\bf 103}
(2009) 202501.
\bibitem{HypX} F.~Cusanno {\it et al.}, Nucl. Phys. A {\bf 835} (2010) 129.
\bibitem{SIMC} R.~Ent {\it et al.}, Phys. Rev. C {\bf 64} (2001) 054610.
\end{thebibliography}

\end{document}